# Continuum surveys with LOFAR and synergy with future large surveys in the 1 – 2 GHz band.

**Raffaella Morganti**[1]
*ASTRON and Kapteyn Institute, University of Groningen, NL*
*E-mail:* morganti@astron.nl

**Huub Rottgering**[2]**, Ignas Snellen, George Miley**
*Leiden Observatory, Leiden University, Postbus 9513, 2300 RA, Leiden, The Netherlands*

**Peter Barthel**
*Kapteyn Institute, University of Groningen, Landleven 12, 9747 AD, Groningen, The Netherlands*

**Philip Best**
*Institute for Astronomy, Royal Observatory, University of Edinburgh, Blackford Hill, EH9 3HJ, UK*

**Marcus Brüggen**
*Jacobs University Bremen, Campus Ring 1, 28759 Bremen, Germany*

**Gianfranco Brunetti**
*INAF - Istituto di Radioastronomia, via Gobetti 101, 40129, Bologna, Italy*

**Krzysztof Chyży**
*Astronomical Observatory of the Jagiellonian University, ul. Orla 171, 30-244 Kraków, Poland*

**John Conway**
*Onsala Space Observatory, SE-439 92 Onsala, Sweden*

**Matt Jarvis**
*Centre for Astrophysics Research, STRI, University of Hertfordshire, Hatfield, AL10 9AB, UK*

**Matt Lehnert**
*GEPI, Observatoire de Paris, CNRS, University Paris Diderot, 5 place Jules Janssen, 92190 Meudon, France*

Radio astronomy is entering the era of large surveys. This paper describes the plans for wide surveys with the LOw Frequency ARray (LOFAR) and their synergy with large surveys at higher frequencies (in particular in the 1 – 2 GHz band) that will be possible using future facilities like Apertif or ASKAP. The LOFAR Survey Key Science Project aims at conducting large-sky surveys at 15, 30, 60, 120 and 200 MHz taking advantage of the wide instantaneous field of view and of the unprecedented sensitivity of this instrument.

[1] Speaker
[2] PI LOFAR Surveys Key Science Project

Four topics have been identified as drivers for these surveys covering the formation of massive galaxies, clusters and black holes using z ≥ 6 radio galaxies as probes, the study of the intercluster magnetic fields using diffuse radio emission and Faraday rotation measures in galaxy clusters as probes and the study of star formation processes in the early Universe using starburst galaxies as probes. The fourth topic is the exploration of new parameter space for serendipitous discovery taking advantage of the new observational spectral window open up by LOFAR.

Here, we briefly discuss the requirements of the proposed surveys to address these (and many others!) topics as well as the synergy with other wide area surveys planned at higher frequencies (and in particular in the 1-2 GHz band) with new radio facilities like ASKAP and Apertif. The complementary information provided by these surveys will be crucial for detailed studies of the spectral shape of a variety of radio sources (down to sub-mJy sources) and for studies of the ISM (in particular HI and OH) in nearby galaxies.

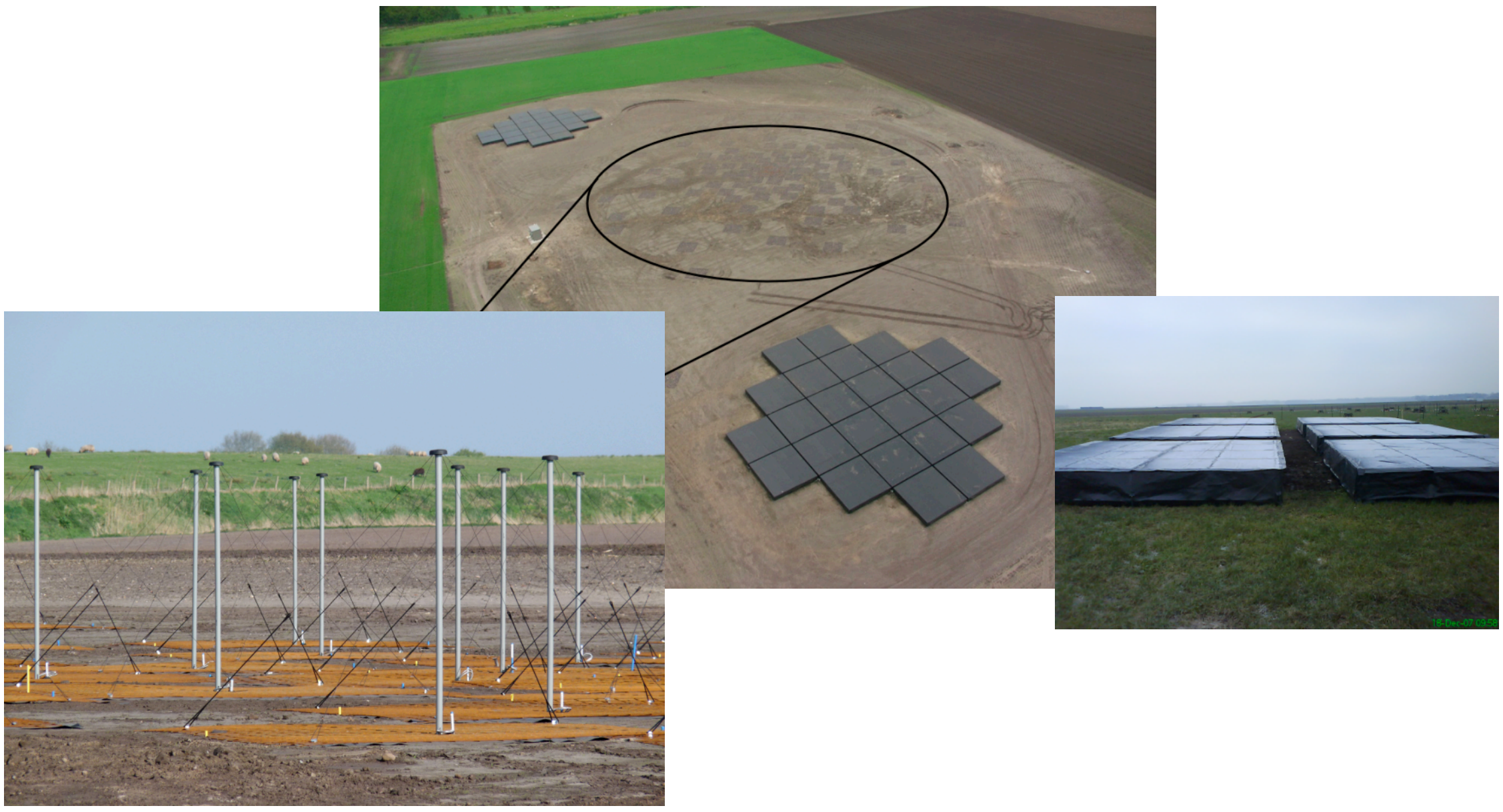

***Fig. 1*** *- The first complete LOFAR station CS 302, showing the Low Band Antennas (enlarged on the left), and the High Band Antenna tiles (also in close up, on the right).*

## 1. Introduction

LOFAR (LOw Frequency ARray) is a new radio telescope designed and built by ASTRON (the Netherlands Institute for Radio Astronomy) in collaboration with Dutch Universities. It has an innovative design that makes use of phased aperture technology, making moving parts redundant, and allowing digital beam forming towards any part of the sky. It also fully exploits recent advances in digital signal processing, fibre-based distributed communication networks and high performance super-computing. LOFAR operates in a largely unexplored region of the electro-magnetic spectrum (from below 20 up to ~240 MHz), and consists of a distributed interferometric array of dipole antenna stations that permit large areas of the sky to be imaged simultaneously. As illustrated in Fig. 1, LOFAR uses two types of antenna: the low-band antenna (LBA, 15-80 MHz) and the high-band antenna (HBA 110-240 MHz), see [3] for a complete system description. Up to 8 (and in the future up to 24) independent fields of view can be instantaneously steered to any position on the sky using electronic beam-forming techniques. Over the next year, at least 36 stations will be built in various locations across the Netherlands. The densely populated core - comprising 18 stations in the inner 2 km - is located in Exloo not far from ASTRON's head-quarters in Dwingeloo. The longest baseline within the Netherlands is ~ 100 km. At least another 8 stations will be located in 4 other European countries – 5 in Germany, and 1 each in Sweden, France & the UK.

## 2. LOFAR Survey Key Science Project

LOFAR is a genuine survey instrument and therefore it is not surprising that one of the Key Science Project (KSP) – the Surveys KSP - has as goal to explore the low-frequency radio sky through several unique surveys. An international survey team (of which the authors of this paper represent the "core" group) is planning to exploit the unprecedented sensitivity and wide instantaneous field of view of LOFAR to conduct large-sky surveys at 15, 30, 60, 120 and 200 MHz. The LOFAR Surveys KSP has from the outset been driven by four key topics. The first three are directly related to the formation of massive black holes, galaxies, and clusters. In particular they deal with the formation of massive galaxies, their central supermassive black holes and their (proto-)cluster environments over a large fraction of cosmic time, from the Epoch of Reionization through to the present time using the radio emission from the AGN. Use diffuse radio emission in galaxy clusters to probe the intracluster magnetic fields as a function of cluster mass and redshift, and finally trace the star-formation history of the Universe through the detection of the radio emission from star-forming galaxies, providing a view of the activity in the Universe free from dust obscuration. The fourth topic is the exploration of new parameter space for serendipitous discovery taking advantage of the new observational spectral window open up by LOFAR. These four key topics drive the areas, depths and frequency coverage of the proposed surveys. In addition to the key themes, the LOFAR surveys will provide a wealth of unique data for a huge number of additional important topics. Furthermore, to maximise the usefulness of the survey data for the Magnetism Key Science Project, the observations will be taken with sufficient bandwidth so that the technique of rotation measure synthesis can be applied to the data. The observations are also planned to be taken in several passes with the individual pointing logarithmically spread in time to facilitate searches for variable sources on various timescales. This will be done in collaboration with members of the Transient Key Science Project.

## 3. Proposed LOFAR surveys

Spread through the first five years of LOFAR operation, the Survey KSP aims at collecting data for three-tier level surveys of the northern sky. Tier 1 is a "*large area*" survey of all northern sky at 15, 30, 60, 120 MHz. The 210 MHz observations are more time consuming and, therefore, full-depth will be limited to smaller area, but all-sky shallower coverage is planned. Tier 2 is a "*deep area*" that aims at reaching a factor few deeper over few x 100 sq. deg at 30,60,120,200 MHz. Finally, Tier 3 is an "*ultra-deep area*" made up of a small number of pointing very deep in one frequency, likely 150 MHz. The expected sensitivities of the proposed surveys are illustrated in Fig. 2.

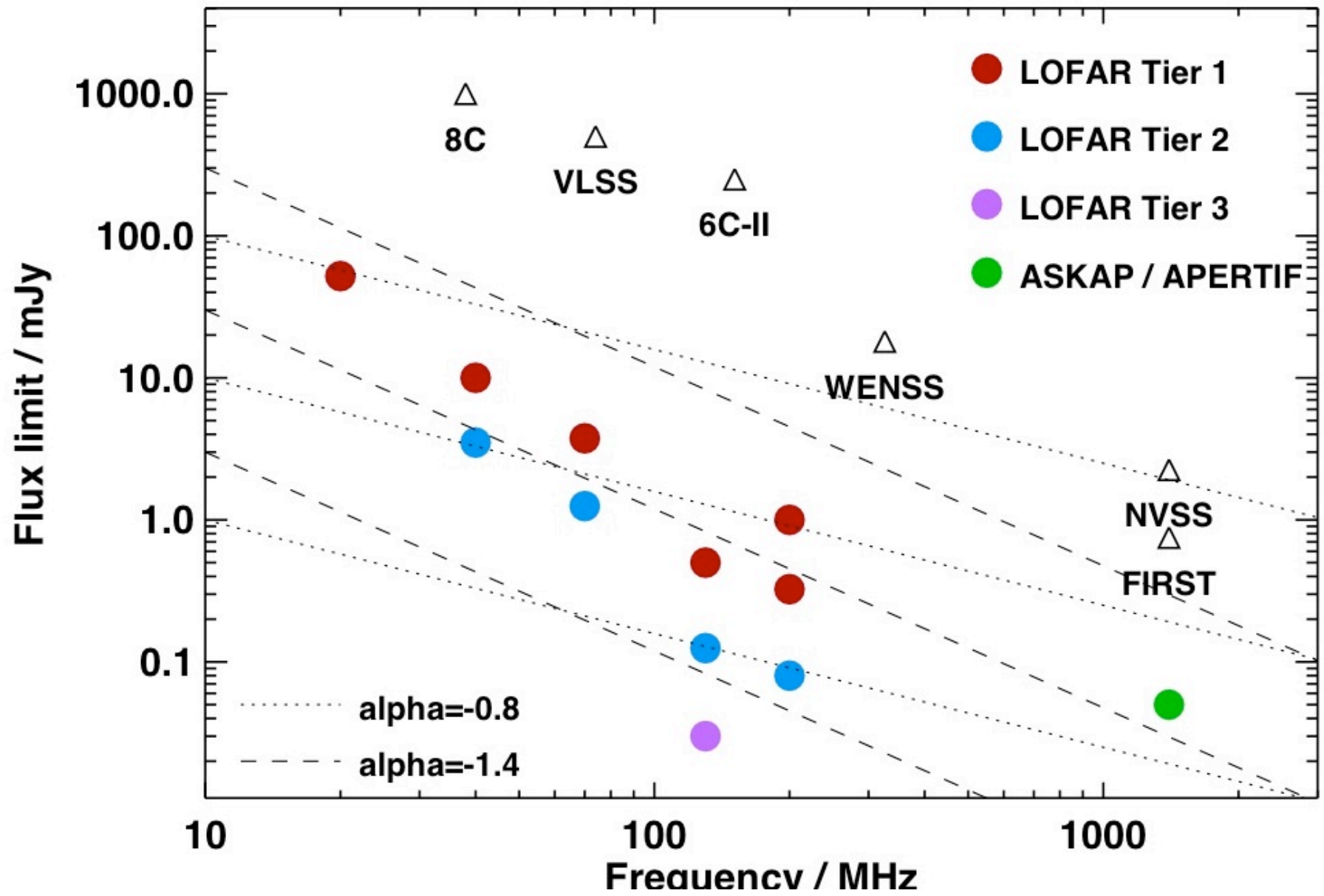

***Fig 2*** *- Flux limits (5 sigma) of the proposed LOFAR surveys compared to other existing radio surveys. The triangles represent existing surveys. The lines represent different power-law ($S \sim \nu^{\alpha}$, with $\alpha = -1.4$ and $-0.8$) to illustrate how, depending on the spectral indices of the sources, the LOFAR will compare to other surveys.*

The large Tier 1 survey is aimed at finding rare objects like high-z radio galaxies and clusters. For example the 120 MHz (2π) survey, with an rms limit of 0.1 mJy/b, should allow us to unveil halo sources in about 500 galaxy clusters, 100 of which beyond z~0.6 (see [4], [2]). Most of radio halos are predicted with very steep spectrum and their discovery will shed new light on the acceleration mechanisms of the relativistic matter in galaxy clusters [1].

The lower frequencies are less sensitive but they also represent an exciting new territory. For example, the 30 MHz 2π-survey will be used to detect high-z sources. The aim is to detect at least 100 Fanaroff-Riley type I and type 2 radio sources at z>7, with S/N>5. Based on the simulations in [18], this requires to reach an rms noise of ~3mJy/b. Both the 30 and 60 MHz surveys will need to be deep enough to detect sources with steep spectrum ($\alpha = -1.6$ with $S \propto \nu^{\alpha}$), among which we will be looking for candidates high redshift objects [11]. The "deep area" aims at obtaining ~25 pointings at each frequency: 30, 60, 120 and 200 MHz. The depth of the 30 and 60 MHz are set to match the depth of the "large area" survey at 120 MHz for typical spectral index sources ($\alpha$=-0.8). The depth of the 120 and 210 MHz are set to detect SF galaxies with 10 $M_{sun}$/yr at z=0.5, and 100 $M_{sun}$/yr at z=2.5

## 4. Synergy between LOFAR and future, large surveys in the 1-2 GHz band

For survey studies, the main advantages of LOFAR are the wide field of view, the different frequencies available and the high spatial resolution that can be reached at such low frequencies. However, most of the work done so far to understand and characterise the various populations of radio sources (e.g. AGN vs starburst, young radio galaxies, faint sub-mJy radio sources etc.) has been done at higher frequencies and in particular around 1.4 GHz. Thus, there is a natural synergy and complementarity between the surveys planned with LOFAR and the surveys in the 1-2 GHz band planned with the wide-field instruments that will be available in the near future. In particular, here we concentrate on facilities like Apertif and ASKAP (see [15] and [6] respectively for a description of these instruments) because of their emphasis on the wide field of view. MeerKAT will be instead particularly interesting for follow up work and observations of relatively small but deeper fields. They all will provide a crucial reference point for LOFAR. Compare to the others, Apertif has the obvious advantage that will be looking at the same sky as LOFAR! The efficiency of LOFAR is going to be lower at lower declinations but hopefully still good enough to allow useful observations that will overlap with e.g. ASKAP surveys. This will also allow follow up observations with other telescopes like VLT and ALMA, crucial for some of the science planned.

There will be at least two aspects for which the combination of the output from LOFAR surveys and ASKAP/Apertif surveys in the 1-2 GHz band will be relevant. It will provide, for a variety of sources, information about the spectral shape over a very wide range of frequencies and it will give the possibility to collect information about the ISM (e.g. presence of HI or OH) for the nearest objects. Fig. 2 includes also the sensitivity obtained for deep fields observed with the WSRT (e.g. HDF and Spitzer First Look, see [7] and [14] respectively) and comparable with what ASKAP will provide. This gives already a first impression on how the different surveys compare.

It is also worth mentioning that both ASKAP and Apertif will actually cover a relatively broad band (likely about 300 MHz, not shown in Fig. 2) that in turn will provide a strong constraint for the spectral index and further help in defining the spectral shape of radio sources.

Young, evolving radio sources (Compact Steep Spectrum – CSS – sources) will be one of the groups of objects particularly interesting for LOFAR. They will allow to constrain evolution models for radio galaxies, the triggering conditions of AGN, to probe the conditions of the ISM in host galaxies and might even point at the first black holes [5]. The turnover in their spectra is known to vary as the source evolves, and being located at lower frequencies for larger sources. Thus, sources with various combination of size/flux will probe different regions of the host galaxies. Particularly interesting and less explored is the kpc region corresponding to the core radii of massive elliptical galaxies. The combination of observations in the 1-2 GHz band and in the frequency range of LOFAR will not only improve our knowledge of the spectral index properties of these sources, but it will also allow to select and study sources peaked in this range of frequencies, i.e. relatively large young sources. For example, it would give the possibility of identify those sources detected at ~1.4GHz but self-absorbed at longer wavelengths i.e. that disappear at (some of) the LOFAR frequencies. The properties and morphologies of these sources, escaping from the more central region of the host galaxies will be very important for understand how radio sources evolve.

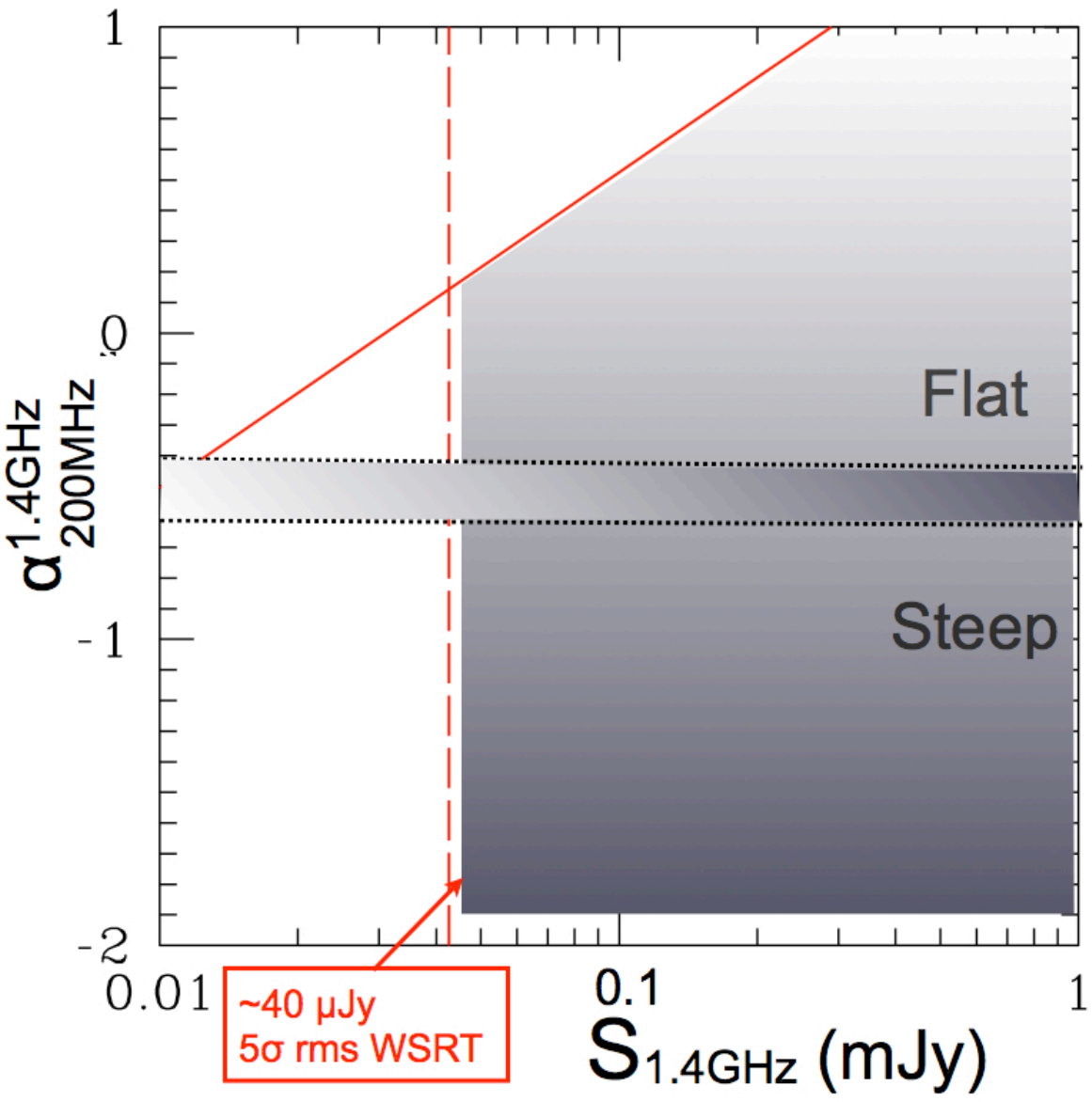

***Fig. 3*** *– Spectral index between 1.4 GHz and 200 MHz as function of the 1.4 GHz flux. The red solid line represents the limit of the spectral index that can be measure between 1.4 GHz data and the fainter sources that we expect to detect in Tier 2 survey. This shows that we will be sensitive not only to steep but also to flat and even inverted source also in the sub-mJy regime.*

Moving to another topic, one of the main questions in the study of deep radio field is to understand the nature of the sub-mJy population of radio sources. Recent studies have shown that even at sub-mJy level the fraction of AGN can be significant (e.g. [16], [8], [17], [9], [18) but it is not clear yet what kinds of AGN dominate at this low flux levels. In particular it is interesting to know whether they are low accretion rate/low efficiency and, therefore, interesting for feedback processes. Some of these studies have suggested that a dominant flat-spectrum or ultra-steep spectrum component is excluded (see [8], [16]) but, on the other hand, flattening of the spectral index has been seen in other studies (see [13]) with no clear explanation of their origin.

Interesting results can be obtained from the combination of higher frequencies (~1.4 GHz) and lower frequencies observations of deep fields. This has been shown in the case of the Lockman Hole by [8] using GMRT 610 MHz data (but limited to 100 arcmin$^2$) in addition to 21cm data from the VLA.

Fig. 2 shows what can be reached by combining LOFAR and wide-field surveys at higher frequencies. Sub-mJy sources and in particular sources with $S_{1.4GHz} \sim 40$ μJy (at the limit of the sensitivity for the WSRT and likely also for ASKAP) will be detected by LOFAR Tier 1 surveys in the high frequency band, if they have spectral index steeper than $\alpha \sim -0.8$, steeper than the typical starforming galaxy. Flat spectrum objects will be visible only if relatively strong in the 1.4 GHz flux (see Fig. 2 for details).

On the other hand, combining 1.4 GHz data with LOFAR Tier 2 and 3 will allow exploring the dependence of the spectrum index with flux as illustrated in Fig. 3. For sources detected at 1.4 GHz down to the μJy level we will be able, for the first time, to study the spectral index including the flat spectrum cases if present. These spectral studies will also provide valuable information for the study of the cosmic star formation history, another main focus of deep surveys around 1.4 GHz as well as LOFAR surveys.

Finally, for low-z objects surveys in the 1-2 GHz band will provide crucial information about the characteristics of their ISM. In particular, it will be possible from these surveys to trace the presence of HI (in emission or absorption) in sources up to $z \lesssim 0.4$ and possibly OH masers in sources up to $z \lesssim 0.7$. This information will be available from the surveys planned by ASKAP and Apertif (see e.g. [10] and [12]). Thus, the input from these HI and OH blind surveys will provide the possibility of studying the gas content for the most nearby radio sources and therefore e.g. the relation between gas and the AGN properties hopefully providing an handle on the duty-cycle activity.

## 5. Conclusions

LOFAR is an ideal telescope for surveys and the KSP Surveys is planning large surveys at various depth and different frequencies (from below 30 up to 200 MHz). These surveys are aimed at finding rare, steep spectrum objects like: high-z radio galaxies, diffuse radio sources in cluster and high-z starforming galaxies. However, they will also enable many other exciting science topics. The synergy with the large surveys in the 1-2 GHz band that will be carried out by telescopes like ASKAP and Apertif is very important. Apart from providing better insight on the spectral characteristics of objects like young, evolving radio sources, faint sub-mJy sources etc. they will also provide crucial information on the properties of the ISM for the most nearby objects in particular through the study of HI and OH.